\documentclass[prb,aps, twocolumn, amsmath,amssymb,superscriptaddress]{revtex4}
\usepackage{float}
\usepackage{amsmath}
\usepackage{amssymb}
\usepackage{amsfonts}
\usepackage{euscript}
\usepackage{enumerate}
\usepackage{hhline}
\usepackage{pslatex}
\usepackage{tabularx}
\usepackage[usenames,dvipsnames]{xcolor}
\usepackage[colorlinks,linkcolor=red,anchorcolor=blue,citecolor=blue]{hyperref}

\usepackage{graphicx}
\usepackage{dcolumn}
\usepackage{bm}
\usepackage[sort&compress]{natbib}
\usepackage{balance}

\makeatletter

\renewcommand{\@biblabel}[1]{#1. }
\renewcommand{\l@figure}[2]{
\@dottedtocline{1}{1.5em}{2em}{Figure #1}{}\vspace{15pt}}

\begin{document}

\title{Dynamic resonance fluorescence in solid-state cavity quantum electrodynamics}

\author{Shunfa Liu}
\thanks{These authors contributed equally}
\affiliation{State Key Laboratory of Optoelectronic Materials and Technologies, School of Physics, School of Electronics and Information Technology, Sun Yat-sen University, Guangzhou 510275, China}

\author{Chris Gustin}
\thanks{These authors contributed equally}
\affiliation{Edward L.\ Ginzton Laboratory, Stanford University, Stanford, California 94305, USA}

\author{Hanqing Liu}
\thanks{These authors contributed equally}
\affiliation{State Key Laboratory for Superlattice and Microstructures, Institute of Semiconductors, Chinese Academy of Sciences, Beijing 100083, China.}
\affiliation{Center of Materials Science and Optoelectronics Engineering, University of Chinese Academy of Sciences, Beijing 100049, China.}

\author{Xueshi Li}
\affiliation{State Key Laboratory of Optoelectronic Materials and Technologies, School of Physics, School of Electronics and Information Technology, Sun Yat-sen University, Guangzhou 510275, China}

\author{Ying Yu}
\affiliation{State Key Laboratory of Optoelectronic Materials and Technologies, School of Physics, School of Electronics and Information Technology, Sun Yat-sen University, Guangzhou 510275, China}

\author{Haiqiao Ni}
\affiliation{State Key Laboratory for Superlattice and Microstructures, Institute of Semiconductors, Chinese Academy of Sciences, Beijing 100083, China.}
\affiliation{Center of Materials Science and Optoelectronics Engineering, University of Chinese Academy of Sciences, Beijing 100049, China.}

\author{Zhichuan Niu}
\affiliation{State Key Laboratory for Superlattice and Microstructures, Institute of Semiconductors, Chinese Academy of Sciences, Beijing 100083, China.}
\affiliation{Center of Materials Science and Optoelectronics Engineering, University of Chinese Academy of Sciences, Beijing 100049, China.}

\author{Stephen Hughes}
\affiliation{\hspace{0pt}Department of Physics, Engineering Physics, and Astronomy, Queen's University, Kingston, Ontario K7L 3N6, Canada\hspace{0pt}}

\author{Xuehua Wang}
\thanks{wangxueh@mail.sysu.edu.cn}
\affiliation{State Key Laboratory of Optoelectronic Materials and Technologies, School of Physics, School of Electronics and Information Technology, Sun Yat-sen University, Guangzhou 510275, China}

\author{Jin Liu}
\thanks{liujin23@mail.sysu.edu.cn}
\affiliation{State Key Laboratory of Optoelectronic Materials and Technologies, School of Physics, School of Electronics and Information Technology, Sun Yat-sen University, Guangzhou 510275, China}
\date{\today}

\begin{abstract}
\noindent \textbf{The coherent interaction between a two-level system and electromagnetic fields serves as a foundation for fundamental quantum physics and modern photonic quantum technology. A profound example is resonance fluorescence, where the non-classical photon emission appears in the form of a Mollow-triplet when a two-level system is continuously driven by a resonant laser. Pushing resonance fluorescence from a static to  dynamic regime by using short optical pulses generates on-demand emissions of highly coherent single photons. Further increasing the driving strength in the dynamical regime enables the pursuit of exotic non-classical light emission in photon number superposition, photon number entanglement, and photon bundle states. However, the long-sought-after spectrum beyond the Mollow-triplet, a characteristic of dynamic resonance fluorescence under strong driving strength, has not been observed yet. Here we report the direct observation and systematic investigations of dynamic resonance fluorescence spectra beyond the Mollow-triplet in a solid-state cavity quantum electrodynamic system. The dynamic resonance fluorescence spectra with up to five pairs of side peaks, excitation detuning induced spectral asymmetry, and cavity filtering effects are observed and quantitatively modeled by a full-quantum model with phonon scattering included. Time-resolved measurements further reveal that the multiple side peaks originate from 
interference of the emission associated with different temporal positions of the excitation pulses. Our work 
facilitates the generation of a variety of exotic quantum states of light with dynamic driving of two-level systems.}
\end{abstract}

\maketitle

Resonance fluorescence (RF) arises from the coherent interaction between 
a single two-level system (TLS) and an optical field, and has played an essential role in the development of quantum optics and its applications in modern photonic quantum technology~\cite{kimble1976theory,loudon2000quantum,fox2006quantum,kimble1977photon}. Despite its conceptual simplicity, it facilitates the exploration of a variety of fascinating physics in the solid-state, such as photon-antibunching~\cite{kim1999single}, Autler-Townes splitting~\cite{xu2007coherent}, Mollow-triplet spectra~\cite{nick2009spin,flagg2009resonantly}, spontaneous emission cancellation~\cite{he2015dynamically} and squeezed light\cite{schulte2015quadrature}. In particular, when a single TLS is resonantly driven by a strong continuous wave (CW) laser, both the ground and excited states are {\it dressed} into two sub-states with a splitting equal to the 
driving Rabi frequency $\Omega$, giving rise to the milestone Mollow-triplet emission spectrum~\cite{astafiev2010resonance,nick2009spin,flagg2009resonantly, ng2022observation}. The characteristic photons associated with the different spectral components of the Mollow-triplet exhibit distinct photon statistics and can be exploited for heralded generations of coherent single photons in the context of photonic quantum technology \cite{aspect1980time,ulhaq2012cascaded}. 

Bringing RF into a {\it dynamic} regime, which requires a sufficiently short pulsed laser, adds unexplored physics to standard quantum optics textbooks and also boosts the development of enabling quantum technologies. To obtain indistinguishable single photons on-demand, single TLSs can be driven by short $\pi$ pulses which deterministically and coherently populate a TLS into the excited state. The successful implementations of $\pi$-pulse resonant excitation on single artificial atoms---epitaxial quantum dots (QDs)---in photonic nanostructures have boosted the development of semiconductor single-photon sources with unprecedented performances in terms of simultaneous high-degrees of source brightness, single-photon purity and photon indistinguishability~\cite{somaschi2016near,unsleber2016highly,he2017deterministic,wang2019towards,uppu2020scalable,tomm2021bright,zhou2023epitaxial}. Very recently, more exotic non-classical states of light were accessed in the high-power driving regime of pulsed RF. At 2$\pi$ and 4$\pi$ pulse areas, two-photon bundle~\cite{fischer2017signatures}, photon number superposition~\cite{loredo2019generation} and even photon number entanglement states~\cite{wein2022photon} were generated from single QDs. For the dynamic RF from QDs, a characteristic spectrum consisting of a central emission line and a multitude of side peaks have been theoretically predicted and quantitatively modeled~\cite{moelbjerg2012resonance,gustin2018spectral}. 

Despite Rabi oscillations of the RF intensity with multiple periods having been steadily achieved in QDs~\cite{fischer2017signatures,liu2018high,loredo2019generation,scholl2020crux,zhai2022quantum}, the direct observation of a dynamic RF spectrum beyond the Mollow-triplet is still illusive to date in any of quantum systems including atoms, semiconductor QDs and superconducting qubits. While some claims of a dynamic Mollow  regime were reported in Ref.~\citenum{fischer2016self}, the excitation conditions were not in a {\it dynamic} regime, which specifically requires a pulse duration that is shorter
than the TLS lifetime  ~\cite{Rzazewski1984Aug,moelbjerg2012resonance,gustin2018spectral}.
To observe such long-sought-after dynamic RF spectra, we apply high-power and short-duration resonant optical pulses to a single QDs deterministically coupled to monolithic semiconductor microcavities.

\begin{center}
	\begin{figure}
		\begin{center}
			\includegraphics[width=1\linewidth]{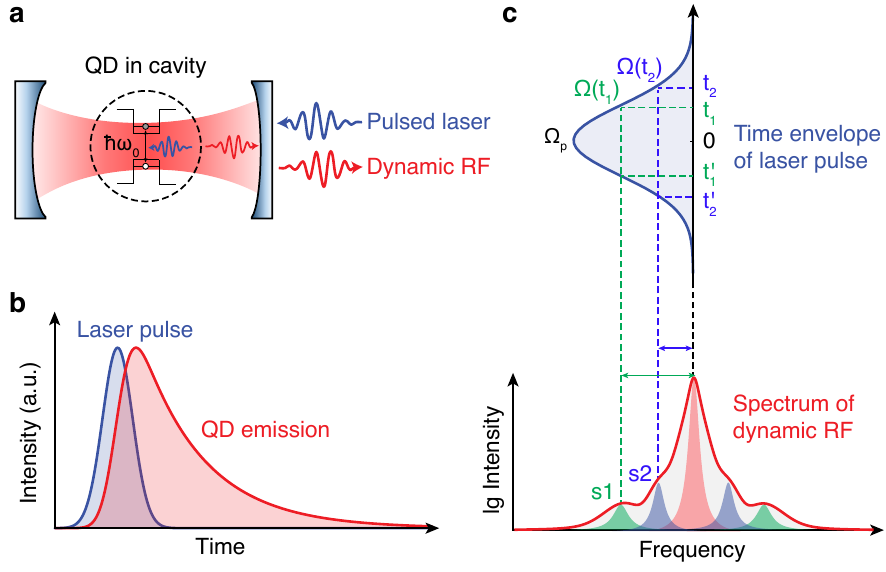}
			\caption{\textbf{Spectrum of dynamic RF with multiple side peaks from a cQED system.} (a) A solid-state cQED system emits dynamic RF under pulsed resonant excitation. (b) Temporal profiles of the excitation pulse and the RF emission. The duration of the laser pulse is shorter than the effective lifetime of the TLS to ensure dynamic interactions. (c) The formation mechanism of the multiple side peaks in the (here resonantly) dynamic RF spectrum. The time-dependent Rabi frequency $\Omega(t_{n})$ is shown as a function of the laser pulse envelope. Each side peak ($s_{1}$, $s_{2}$) corresponds to two specific temporal positions ($t_{1}$ and $t_{1}^{\prime}$, $t_{2}$ and $t_{2}^{\prime}$) in the driving pulse between where constructive interference occurs. }
			\label{fig:Fig1}
		\end{center}
	\end{figure}
\end{center}

\section{Principles of dynamic RF}

As shown in Fig.~\ref{fig:Fig1}(a), we excite a semiconductor artificial atom (QD) by using resonant optical pulses from one side of the cavity. The dynamic RF is emitted towards the same side of the cavity and the reflected laser is suppressed via a cross-polarization technique~\cite{he2013demand}. To ensure the RF is in the {\it dynamic} regime, the duration of laser pulses should be shorter than the effective lifetime of the QDs~\cite{moelbjerg2012resonance}, as schematically shown in Fig.~\ref{fig:Fig1}(b). It has been predicted that the spectrum of dynamic RF consists of multiple side peaks, which is in stark contrast to the conventional Mollow-triplet under CW excitations~\cite{moelbjerg2012resonance,gustin2018spectral}. The physical mechanism of those side peaks in the dynamic RF spectrum is schematically shown in Fig.~\ref{fig:Fig1}(c). Driven by strong optical short pulses, the QD is expected to exhibit a set of continuous Mollow-triplets, each with an instantaneous Rabi frequency $\Omega(t_{n})$ corresponding to certain temporal positions $t_{n}$ in the pulses. For the Rabi frequencies satisfying a specific phase relation~\cite{moelbjerg2012resonance} [e.g., $t_{1}$ and $t_{2}$ in Fig.~\ref{fig:Fig1}(c)],  constructive interference of the emission between two temporal positions [here between $t_{1}$ and $t_{1}^{\prime}$, $t_{2}$ and $t_{2}^{\prime}$ ] of the pulses lead to the formation of discrete side peaks in the emission spectrum~\cite{moelbjerg2012resonance}. 

Our experiments are described by a full-quantum model for the light-matter interactions. Specifically, 
the cavity quantum electrodynamic (cQED) system under pulsed driving is modeled by a quantum master equation, in a frame rotating at the laser frequency, as a time-dependent Lindblad master equation (with $\hbar=1$)
\begin{equation}\label{eq:1}
	\dot{\rho} = -i[H(t),\rho] + \frac{\kappa}{2}\mathcal{L}[a]\rho + \frac{\gamma}{2}\mathcal{L}[\sigma^-]\rho,
\end{equation}
where the system Hamiltonian $H(t)$ is 
\begin{equation}
	H(t) =   \Delta_c a^{\dagger} a + \Delta_x\sigma^+\sigma^- +g(a\sigma^+ + a^{\dagger} \sigma^-) + H_{\rm drive}(t),
\end{equation}
where $a,a^\dagger$
are the annihilation/creation operators for the cavity mode,
and $\sigma^+,\sigma^-$ are the Pauli
operators of the TLS.
The detunings are $\Delta_c = \omega_c - \omega_{\rm L}$, $\Delta_x = \omega_x - \omega_{\rm L}$, where $\omega_c$, $\omega_{\rm L}$, and $\omega_x$ are the frequencies of the cavity resonance, driving laser, and exciton transition, respectively. 
To model dissipation, $\kappa$ and $\gamma$ are the decay rates for the cavity and TLS, respectively, and $\mathcal{L}[A]\rho = 2A\rho A^{\dagger} - A^{\dagger}A \rho - \rho A^{\dagger} A$ is the Lindblad superoperator. The term $g$ is the effective cavity coupling rate, much smaller than $\kappa$ which ensures weak coupling.
For the driving term, we employ a cavity drive $H_{\rm drive}(t) = \frac{\Omega(t)}{2}(a+a^{\dagger})$,
with Rabi frequency $\Omega(t)$, to model the cQED system, which is different from the previous theoretical works~\cite{moelbjerg2012resonance, gustin2018spectral} but necessary to describe our cQED systems~\cite{SI} (see SI). In addition, we include a time-dependent polaron-transform master equation model of acoustic phonon coupling, which is well-established in describing optically driven QD exciton dynamics~\cite{Gustin2018Jul,Nazir2016Feb}, and we fit the phonon coupling strength to experiment using Rabi oscillation data (see SI). The output observable time-dependent incoherent spectra, $S(\omega,t)$, is obtained from
\begin{equation}
	S(\omega,t) = \text{Re}\left[ \int_0^{t} dt' \int_0^{t-t'} d\tau \langle a^{\dagger}_{\delta}(t')a_{\delta}(t'+\tau)\rangle e^{i(\omega-\omega_{\rm L})\tau} \right],
\end{equation}
where $a_{\delta} = a - \langle a \rangle$, and the long-time spectra is $S(\omega) = \lim_{t\rightarrow\infty}S(\omega,t)$.

\begin{figure*}
	\includegraphics[width=1.0\linewidth]{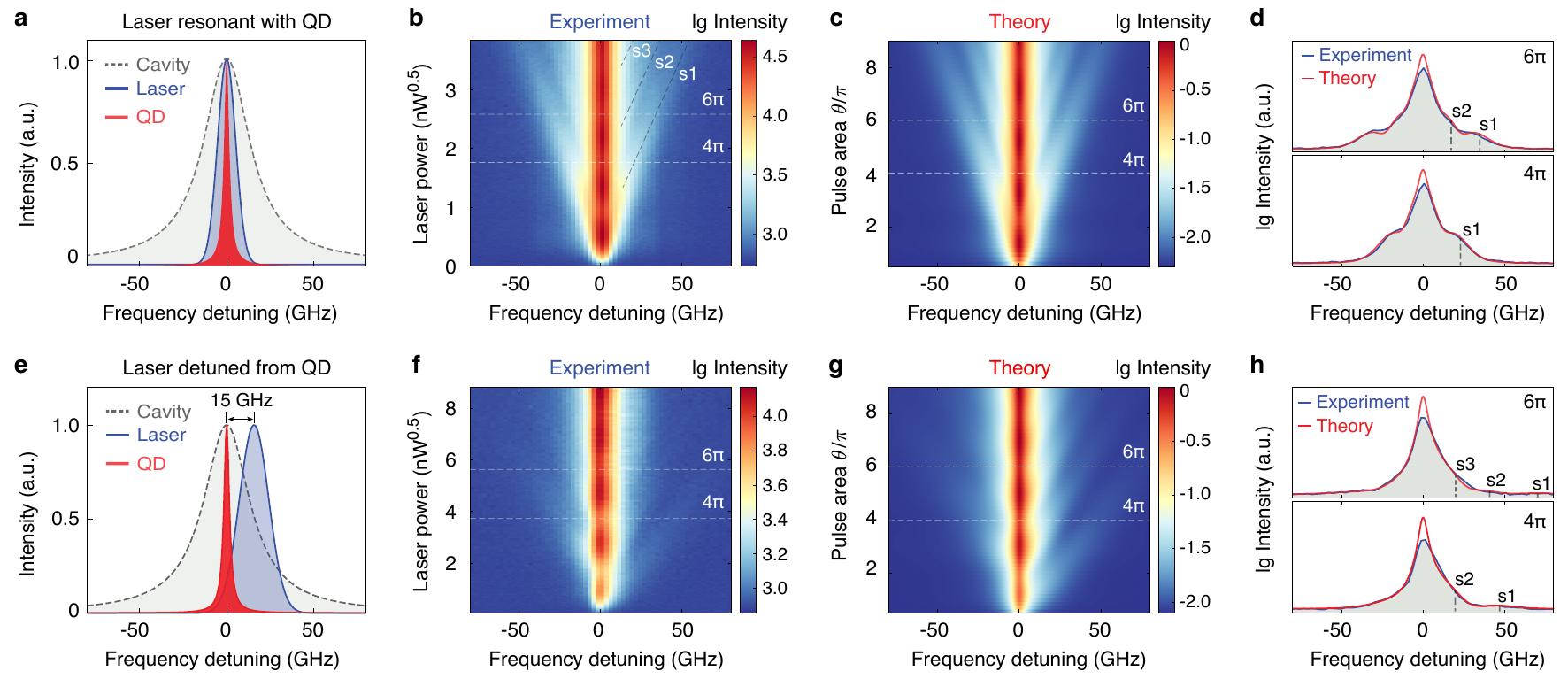}
	\caption{\textbf{Dynamic RF spectra beyond the conventional Mollow-triplet.} (a) Resonant excitation condition at 14~K: $\Delta_c=0$ and $\Delta_x=0$, with a pulse width of 54 ps. (b) Contour plot of the measured RF spectra [under the condition in (a)] as a function of the driving pulse amplitude. The side peaks are labeled with s1, s2, and s3. (c) Simulated dynamic RF spectra for the experiments in (b). (d) Comparisons of the experimental and calculated spectra [under the condition in (a)] at 4$\pi$ and 6$\pi$ pulse areas. (e) Off-resonant excitation condition at 14~K: $\Delta_c=0$ and $\Delta_x/(2\pi) =-15~\rm GHz$, with a pulse width of 24 ps. (f) Contour plot of the measured RF spectra [under the condition in (e)]  as a function of driving pulse amplitude. (g) Simulated pulsed RF spectra for the experiments in (f). (h) Comparisons of the experimental and calculated spectra 
 [under the condition in (e)] at 4$\pi$ and 6$\pi$ pulse areas.}
	\label{fig:Fig2}
\end{figure*}

\section{Observation of the dynamic RF}

Experimentally, the pulsed resonant excitations are implemented in a solid-state cQED system consisting of a single QD deterministically coupled to a semiconductor micropillar cavity~\cite{su2018bright,liu2021dual,wei2022tailoring}. The characterization setup is schematically shown in Fig.~\ref{fig:Fig1E}. Dynamic RF spectra beyond Mollow-triplet are universally observed in our samples. Here, we present experimental results from three representative devices with varied Q-factors, where Q is the cavity mode quality factor (see 
Extended Data
Fig.~\ref{fig:Fig2E}), and all of the measurements are based on charged exciton states ($\rm X^+$) of the QDs for avoiding the influence of fine structure splittings of neutral exciton states.
For device 1, featuring a Q-factor of 10584 and a Purcell factor of 8.25 (on-resonant QD lifetime of 66.3~ps), 54 ps pulses are employed to drive the system under the $\Delta_c =\Delta_x=0$ condition at a temperature of 14 K, as illustrated in Fig.~\ref{fig:Fig2}(a). 

Dynamic RF spectra in the experiment are collected by using an integration time of only 0.1~s without any background correction process. Such a high signal-to-noise ratio greatly benefits from the cQED effect in our system and would not be possible by using QDs in bulk. The evolution of dynamic RF spectra as a function of the pulse amplitude is plotted in Fig.~\ref{fig:Fig2}(b). With increased driving strength, Rabi oscillations up to areas of 9$\pi$ are clearly observed for the central emission peak. Notably, new emission peaks emerge from the central resonance frequency at certain excitation powers and move outwards. This unique multi-peak behavior is beyond the conventional dressed state description and can be well reproduced by our full-quantum model with phonon-induced decoherence included, as presented in Fig.~\ref{fig:Fig2}(c). In Fig.~\ref{fig:Fig2}(d), we obtain excellent agreements between the experiment and theory for the dynamic RF spectra under pulse areas of 4$\pi$ and 6$\pi$. In contrast to the static counterpart, where the CW RF spectra exhibit a Mollow-triplet consisting of three peaks, pulsed excitation gives rise to several side peaks which are all spectrally closer to the central peak and lower in amplitude than the Mollow side peaks would. Another pronounced difference between the static and dynamic RF is the spectral asymmetry under detuned excitations~\cite{gustin2018spectral}. Under a detuned CW drive, satellite sidebands always remain spectrally symmetric with respect to the central peak in the absence of pure dephasing. On the contrary, the multi-peak structure of the dynamic RF spectrum features a pronounced off-resonant spectral asymmetry under detuned pulsed excitation, which is attributed to adiabatic excitation of the quasi-dressed states of the time-dependent system~\cite{gustin2018spectral}. Under a detuning of $\Delta_x/(2\pi) =-15~\rm GHz$ (Fig.~\ref{fig:Fig2}(e)), the spectra of dynamic RF show appreciable asymmetries of the side peaks in both the spectral position and amplitude (Fig.~\ref{fig:Fig2}(f)), which are in excellent 
agreements with the modelling (Fig.~\ref{fig:Fig2}(g,~h)) and prior predictions~\cite{gustin2018spectral}. We further present the dynamic RF spectra at 2$\pi$ and 4$\pi$ as a function of the laser detuning in Fig.~\ref{fig:Fig3E}. The influence of the pulse duration on the dynamic RF spectra is investigated in Fig.~\ref{fig:Fig4E} and Fig.~\ref{fig:Fig5E}.

\begin{figure*}
	\begin{center}
		\includegraphics[width=1.0\linewidth]{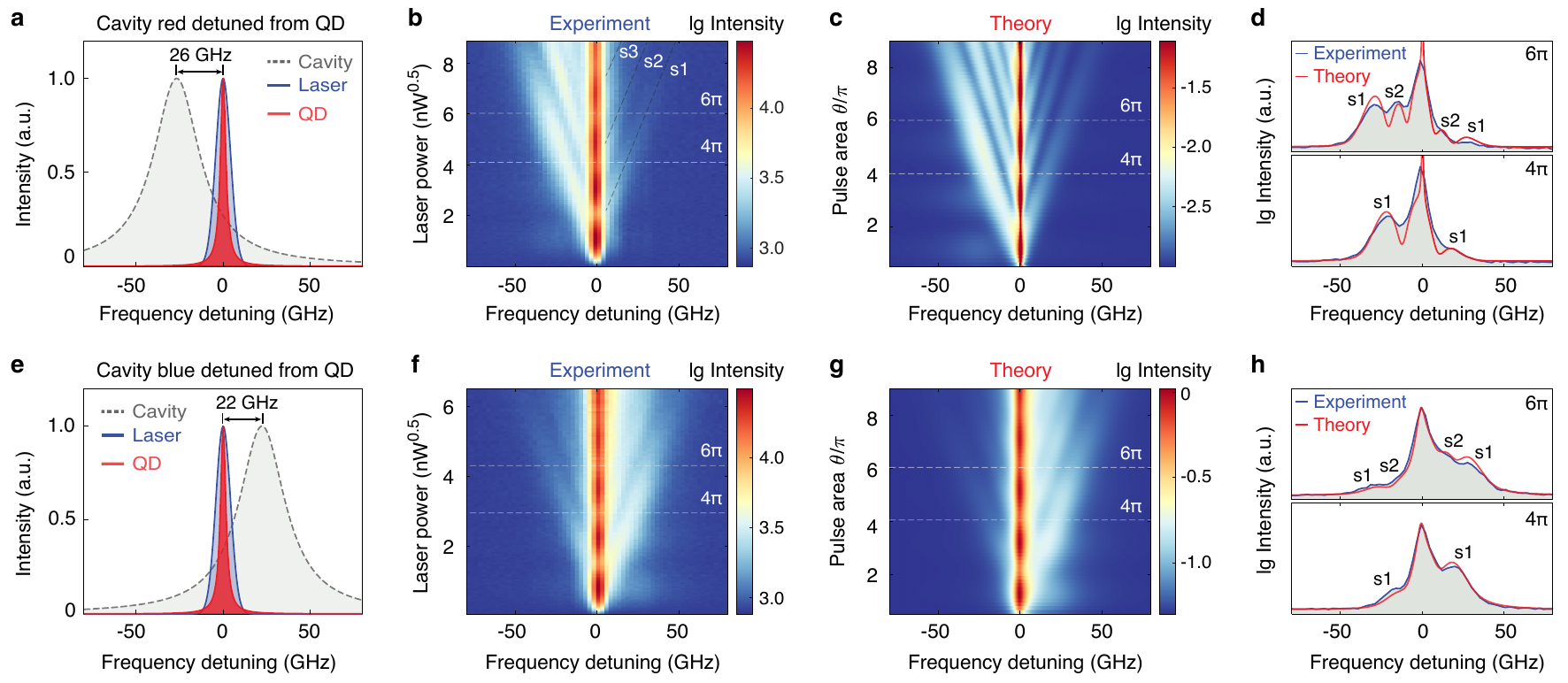}
		\caption{\textbf{Cavity-enhanced side peaks in the spectra of dynamic RF.} (a) Cavity blue-detuned condition at 4~K: $\Delta_c/(2\pi) =-26~\rm GHz$ and $\Delta_x=0$, with a pulse width of 54 ps. (b) Contour plot of the measured RF spectra [under the condition in (a)] as a function of drive pulse amplitude. (c). Simulated dynamic RF spectra for the experiments in (b). (d) Comparisons of the experimental and calculated spectra [under the condition in (a)] at 4$\pi$ and 6$\pi$ pulse areas. (e) Cavity red-detuned condition at 19~K: $\Delta_c/(2\pi) =22~\rm GHz$ and $\Delta_x=0$, with a pulse width of 54 ps. (f) Contour plot of the measured RF spectra (under the condition in (e)) as a function of drive pulse amplitude. (g) Simulated pulsed RF spectra for the experiments in (f). (h) Comparisons of the experimental and calculated spectra [under the condition in (e)] at 4$\pi$ and 6$\pi$ pulse areas.}
		\label{fig:Fig3}
	\end{center}
\end{figure*}

 \begin{center}
	\begin{figure*}
		\begin{center}
			\includegraphics[width=0.85\linewidth]{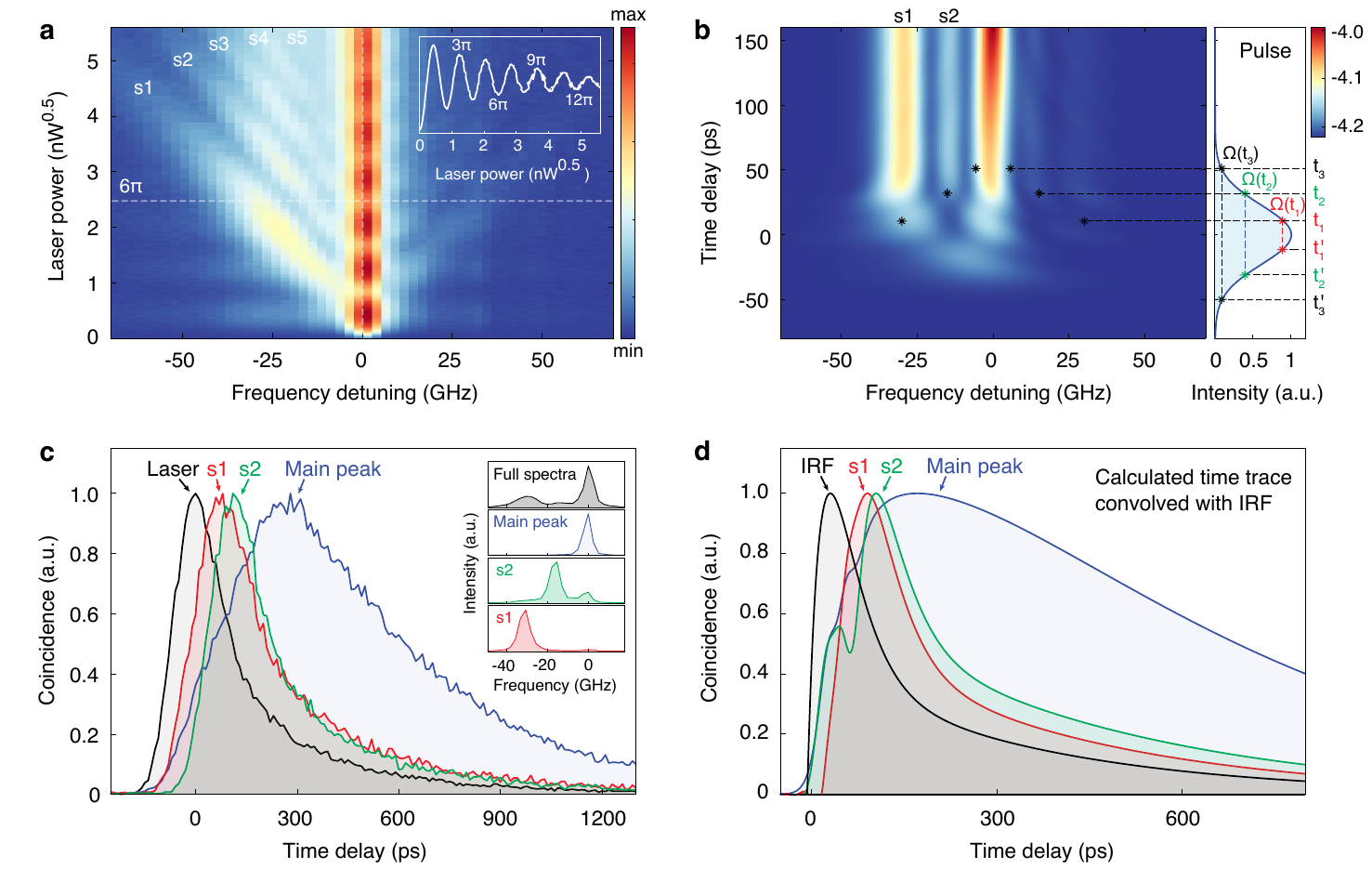}
			\caption{\textbf{Time ordering of the side peaks in the spectra of dynamic RF.} (a) Contour plots of the measured RF spectra of device 2 as a function of driving pulse amplitude, under the condition of $\Delta_c=0$ and $\Delta_x/(2\pi) =-29$ GHz, with a pulse width of 54 ps. Due to the enhancement of the cavity mode, 5 pairs of side peaks (labeled with s1 to s5) with intensities in the same order as the main peak are clearly observed. Insert: photon count rate of the main peak as a function of the driving pulse amplitude, presenting a distinct Rabi oscillation with up to 14$\pi$ periods. (b) Calculation of the time-dependent spectra of the dynamic RF in device 2.  Also shown is the time-dependent Rabi frequency $\Omega(t_{n})$. The time points satisfying the constructive interference condition are indicated with marks and connected with vertical lines. (c) Time-resolved measurement of the individual side peak at a pulse area of 6$\pi$. Insert: full and filtered spectra of the dynamic RF at 6$\pi$. The two side peaks shaded with green and red as well as the main peak (shaded with blue) are spectrally filtered for the lifetime measurement. (d) Calculated time traces of filtered peaks which have been convoluted with the IRF of the APD.}
			\label{fig:Fig4}
		\end{center}
	\end{figure*}
\end{center}

To further visualize the multiple side peaks in the dynamic RF spectra, we intentionally move the cavity mode from the central emission line towards the side peaks via temperature tuning. Under a red detuning of $\Delta_c/(2\pi) =-26~\rm GHz$ at 4 K, the intensities of the on-resonant side peaks are significantly enhanced while their spectral positions remain essentially unchanged. On the other side of the central emission peak, the emission intensities of the off-resonant side peaks are greatly suppressed, as experimentally shown in Fig.~\ref{fig:Fig3}(b). The dynamic RF spectra on the cavity-detuned condition can be quantitatively reproduced with the full-quantum model in Fig.~\ref{fig:Fig3}(c.d). For comparison, a similar cavity filtering effect is also observed in the blue-detuned condition of $\Delta_c/(2\pi) =22~ \rm GHz$ at $ 19~\rm K$, as presented in Fig.~\ref{fig:Fig3}(e-h). As seen, the blue-detuned side peaks are narrower while the red-detuned emission tends to smear out. Such a difference is attributed to the distinct decoherence rates of acoustic phonons at different temperatures, as well as the asymmetric preferential nature of phonon emission-assisted processes over phonon absorption-assisted processes at low temperatures, which is also taken into account rigorously in our model. This is a unique feature of our solid-state TLS, and further highlights
the rich physics of our cQED system. We noted all the features in the dynamic RF spectra can only be captured by our full-quantum model which includes cavity excitation and phonon scattering, as presented in Fig.~\ref{fig:Fig6E}.

\section{Time ordering of the side peaks in dynamic RF}
The appearance of a side peak in the spectra of dynamic RF has been suggested as a consequence of the  constructive quantum interference of the emissions associated with the different temporal positions of the excitation pulse\cite{moelbjerg2012resonance}. For device 2, with a Q-factor of 14,652 and a Purcell factor of 10.4 (on-resonance lifetime of 61.3~ps), the cavity is red-detuned from the QD with $\Delta_x/(2\pi)  = - 29\ \rm GHz$ at 4 K. As shown in Fig.~\ref{fig:Fig4}(a), since the dephasing caused by phonons is minimized at a lower temperature, a Rabi oscillation of the main peak up to 14$\pi$ is clearly observed. Moreover, thanks to the enhancement of the cavity mode, the side peaks are bright enough for time-resolved measurement after spectral filtering. 

In the calculated time-dependent spectra of Fig.~\ref{fig:Fig4}(b), the outer peak $s_{1}$ in the spectra corresponds to the temporal position $t_{1}$ close to the pulse center and appears earlier than the inner peak $s_{2}$ linked to the temporal position $t_{2}$ far from the pulse center. This is because the quantum interference between $t_{1}$ and $t_{1}^{\prime}$ happens earlier than that between $t_{2}$ and $t_{2}^{\prime}$ within the excitaion pulses. Such a specific time ordering of the side peaks can be probed by time-resolved measurements. We spectrally filter the side peaks (at a 6$\pi$ pulse area) and measure the lifetime of each peak, as presented in Fig.~\ref{fig:Fig4}(c). A time ordering between different side peaks is clearly resolved, which directly confirms that the constructive interference of the emission at different pulse positions contributes to the formation of the spectra beyond the Mollow-triplet. The experimental observation agrees well with the calculated time trace of each filtered peak, when convoluted with the instrument response function (IRF) of the avalanche photon diode (APD), as presented in Fig.\ref{fig:Fig4}(d).

 \begin{center}
	\begin{figure*}
		\begin{center}
			\includegraphics[width=1.0\linewidth]{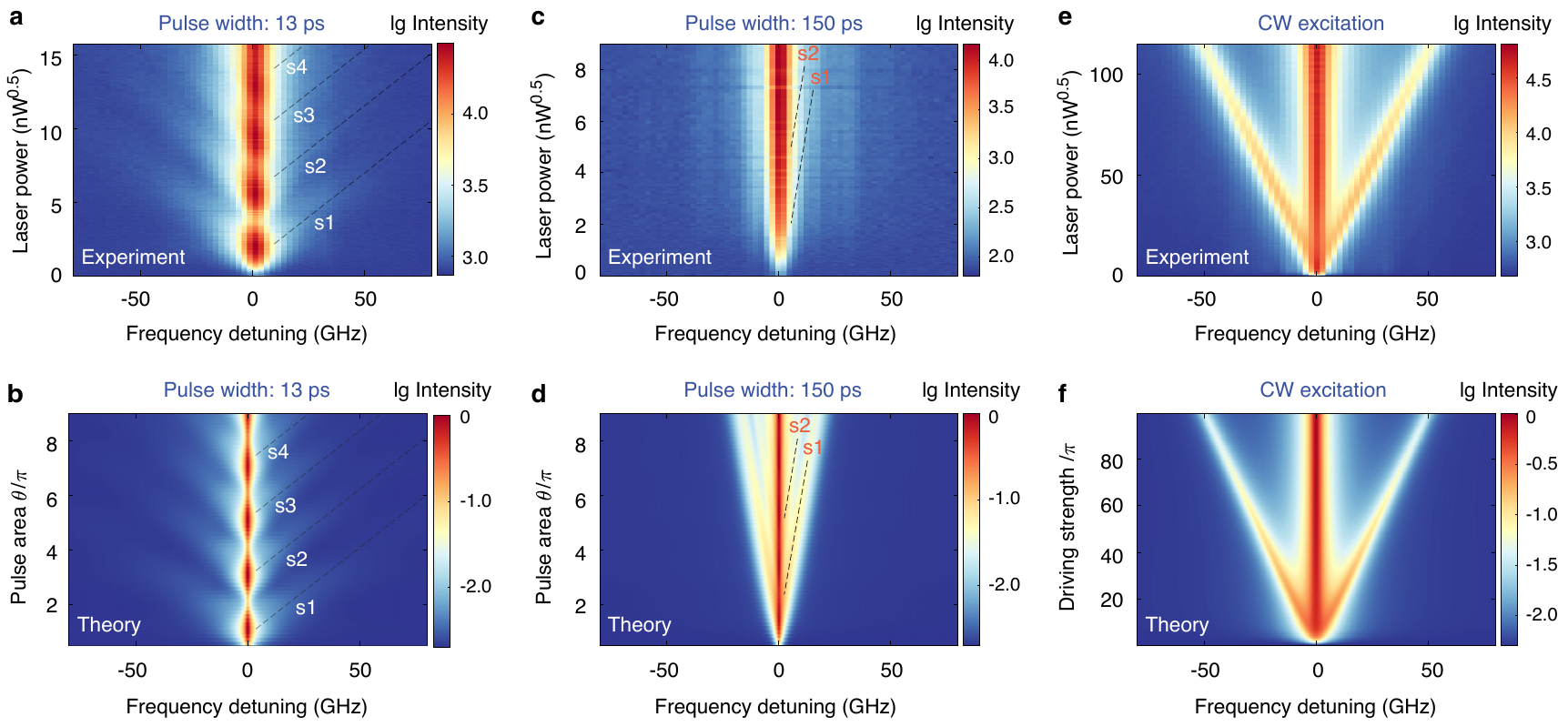}
			\caption{\textbf{Transition from pulsed RF to static RF.} (a) Experiment and (b) theory of the dynamic RF with pulse duration shorter than the effective  QD lifetime. (c) Experiment and (d) theory of the quasi-static RF with pulse duration twice of the effective QD lifetime. (e) Experiment and (f) theory of the static RF with CW excitation.}
			\label{fig:Fig5}
		\end{center}
	\end{figure*}
\end{center}

\section{Transition from dynamic to static RF}

Finally, we investigate the transition from the dynamic RF spectra with multiple side peaks to the well-known Mollow-triplet spectra by increasing the duration of the optical pulses. For device 3, with a Q-factor of 6796 and a Purcell factor of 7 (on-resonant QD lifetime of 70~ps), up to four pairs of side peaks are observed by using 13~ps driving pulses, as experimentally shown in Fig.~\ref{fig:Fig5}(a) and numerically calculated in Fig.~\ref{fig:Fig5}(b). Moving to the quasi-static regime with a pulse duration of 150~ps (more than twice of the QD lifetime), the side peaks get close to each other due to the relatively small effective Rabi frequency $\Omega$ for longer pulses with the same pulse area. Therefore, it becomes difficult to spectrally resolve the multiple side peaks both in the experiment (Fig.~\ref{fig:Fig5}(c)) and theory (Fig.~\ref{fig:Fig5}(d)). We note that 
very similar Mollow-triplet-like spectra of quasi-CW RF have previously been studied in a strongly-coupled QD-cavity system via a self-homodyning interferometric technique~\cite{fischer2016self}. Under the CW excitation condition, the expected Mollow-triplet spectra are recovered, as present in Fig.~\ref{fig:Fig5}(e,f). We point out that the Rabi splitting in our Mollow-triplet spectra is up to 50~GHz, which is remarkable and can be attributed to the cavity-enhanced driving field.

\section{Conclusions}

We have demonstrated the long-sought-after spectra of dynamic RF beyond the conventional Mollow-triplet in a solid-state cQED system. The unique features of multiple side peaks, excitation-induced spectral asymmetry, and cavity filtering effects are quantitatively reproduced by a full quantum model with phonon-induced decoherence included. The time-ordering of individual side peaks is directly observed, suggesting that their formation is due to the constructive interference of the emissions belonging to different temporal positions of the driving pulses. This achievement may bring unprecedented opportunities in the generations of N-photon bundles, superposition of photon number state, and photon number entangled states via dynamic RF at a high-power excitation level. Looking to future prospects, it is exciting to explore the quantum nature of the photonic bound state in the dynamic RF~\cite{tomm2023photon}. From a device point of view, multi-color triggered single-photon sources with spectral tunability could be pursued by multiplexing the side peaks in the spectra of the dynamic RF~\cite{he2015dynamically}. Our work serves as an important breakthrough toward the development of solid-state photonic quantum technology.


\vspace{0.8em}\noindent  \textbf{Data availability}

\noindent {The data that support the findings of this study are available within the paper and the Supplementary Information. Source data and other relevant data are available from the corresponding authors upon reasonable request.}

\vspace{0.8em}\noindent \textbf{Acknowledgements}

\noindent {This research was supported by the National Key Research and Development Program of China (2021YFA1400800, 2018YFA0306101); the National Natural Science Foundation of China (62035017); the Guangdong Special Support Program (2019JC05X397); the National Super-Computer Center in Guangzhou;
and the Natural Sciences and Engineering Research Council of Canada (NSERC). 
}

\vspace{0.8em}\noindent \textbf{Author Contributions}

\noindent J.~L. conceived the project; S.~F.~L., J.~L. and H.~Q.~L. designed the epitaxial structure and the devices; H.~Q.~L., Y.~Y., and H.~Q.~N. grew the quantum dot wafers; S.~F.~L. and X.~S.~L. fabricated the devices; S.~F.~L. built the setup and performed the optical measurements; C.~G. performed all the theoretical modeling with input from S.~H., and wrote the main theory parts of the paper and the SI; S.~F.~L., J.~L., and C.~G. analyzed the data; J.~L. and S.~F.~L. prepared the main manuscript with inputs from all authors; J.~L., S.~H., Z.~C.~N. and X.~H.~W. supervised the project.

\noindent \textbf{Conflict of Interest}

\noindent The authors declare no conflict of interest.


\section{methods}
\noindent \textbf{Sample fabrication:}
Two III-V wafers consist of a single layer of low density In(Ga)As QDs between the top(bottom)  GaAs/$\rm{Al_{0.9}Ga_{0.1}As}$ distributed Bragg reflector (DBRs) have been grown via molecular beam epitaxy (MBE) for this study. Device 1 and device 2 are from wafer A with 20(30) top(bottom) DBR pairs, and device 3 is from wafer B with 18(28) top(bottom) DBR pairs. The deterministically coupled QD-in-micropillars are fabricated using a wide-field fluorescence imaging technique. First, metallic alignment marks are formed on the surface of the wafer by using E-beam lithography, metal evaporation, and standard lift-off processes. Then the wide-field optical images containing both alignment marks and the QDs are obtained by a bi-chromatic illumination technique. The positions of the QDs respective to the alignment marks are extracted from the fluorescence images. Finally, a large number of micropillars containing single QDs are realized through the second E-beam lithography followed by a chlorine-based dry etch process. Devices with $\rm X^+$ state near-resonance with the fundamental cavity mode of the micropillar are chosen for further measurements.

\noindent \textbf{Optical characterizations:}
The schematic of the setup for optical characterizations is presented in Extended Data Fig.~\ref{fig:Fig1E}. The sample is located in a close-circle\ cryostat with a base temperature of 3.5 K. A Femtosecond (fs) pulse from a Ti-sapphire oscillator is shaped to picosecond (ps) pulses for resonant excitation by a home-made 4f system and gratings. The central wavelength of the ps pulse is tunable, and the pulse width can be changed from 10 ps to 55 ps. For generating 150 ps pulses, a fiber-coupled and EOM-controlled lithium niobate Mach–Zehnder (MZ) interferometer was used to modulate a tunable CW laser. The shaped pulse is then sent into a customized confocal microscope with polarization controllers in both the excitation and collection paths. A white light lamp in the cryogenic confocal system is used as an illumination source for imaging. Meanwhile, it reduces the charge noise around the QD for improving the quantum efficiency (QE) of the QD under resonant excitation. The emitted single photons are collected by a single-mode fiber and sent to a spectrometer for spectral analysis. The  spectrometer is equipped with a $1800~\rm l/mm$ grating, a focal length of 750 mm, and features a spectra resolution of $\sim8$ GHz at 900 nm. A homemade etalon filter with a bandwidth of $\sim$8 GHz is employed to filter out each peak from dynamic RF spectra for the lifetime measurement. A fast APD with a combined fast response of 30 ps and slow response of 350 ps is used for time-resolved measurement.

\noindent \textbf{Theory modeling:}
We use a time-dependent polaron-transformed master equation approach with a quantized cavity mode to simulate the pulsed RF experiments. We exploit a time-dependent coherent displacement transformation to efficiently model the cavity driving as an effective TLS drive, and spectra are calculated using the quantum regression theorem for two-time correlation functions. We fit the phonon coupling strength to Rabi oscillation measurements. See SI for further details.

\clearpage
\newpage
\onecolumngrid \bigskip

\begin{center} {{\bf \large EXTENDED DATA}}\end{center}

\setcounter{figure}{0}
\makeatletter
\renewcommand{\thefigure}{E\@arabic\c@figure}

\begin{center}
	\begin{figure}[!h]
		\includegraphics[width=1\linewidth]{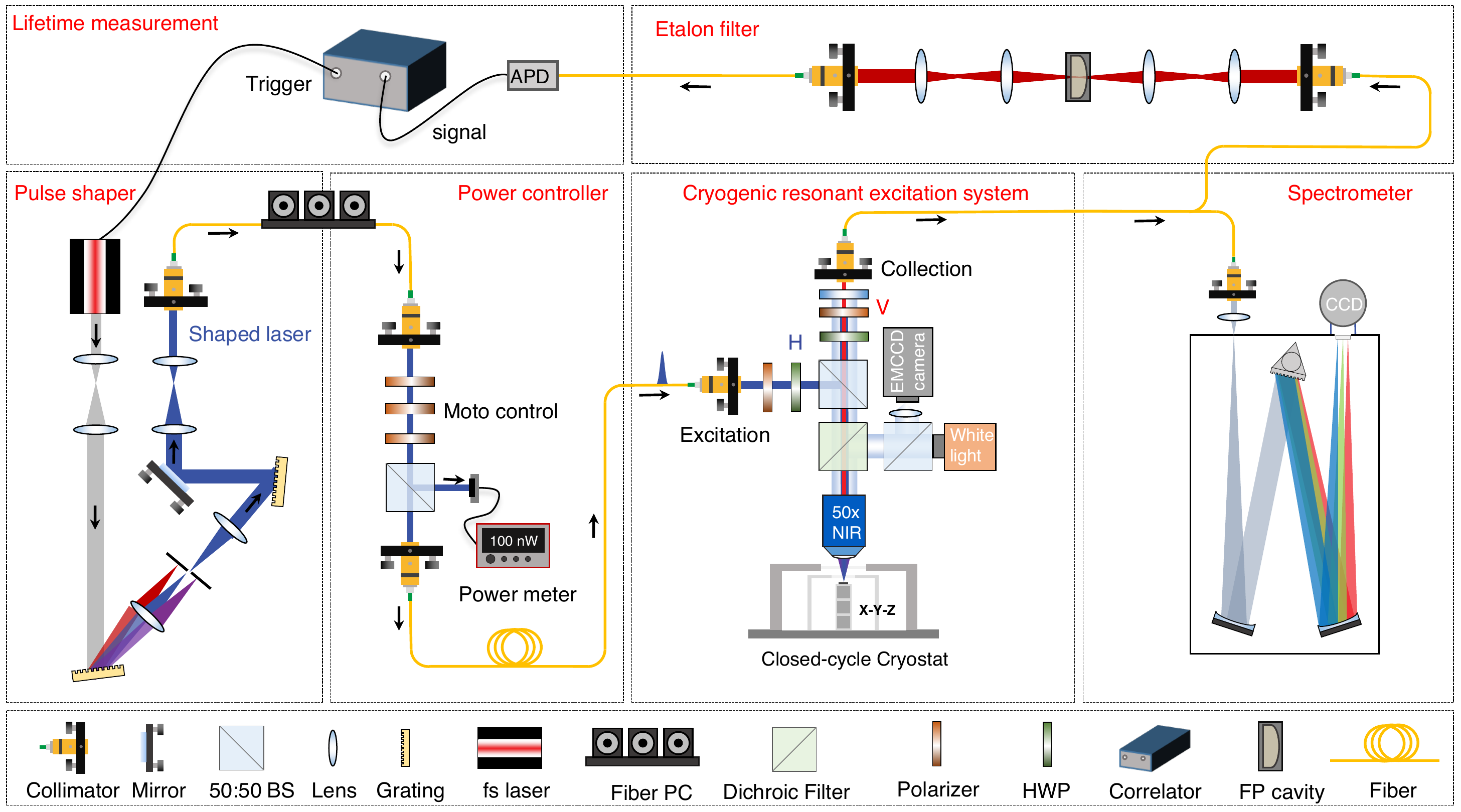}
		\caption{\textbf{Schematic of the setup for optical characterizations.} The setup consists of 6 functional sections including a cryogenic resonant excitation system, a pulse shaper, a power controller, a spectrometer, an etalon filter, and a lifetime measurement section. The measurements start from shaping the femtosecond (fs) pulses to a picosecond (ps) pulse with tunable wavelength and controllable pulse width between 10 ps and 55 ps by a homemade 4f system and gratings. The power of the shaped ps pulse is then controlled by rotating a polarizer between two fixed polarizers with the same polarization axis and monitored by a power meter. The resonant excitation of the QD is based on a cross-polarization scheme, pairs of the polarizer and half-wave plate are inserted into the excitation and collection optical path for polarization control. RF from QD is coupled to an optical fiber, guided to a spectrometer for spectrum analysis, or used for time-resolved measurement after filtering with an etalon. BS: beam splitter, HWP: half-wave plate, LP filter, PC: polarization controller.}
		\label{fig:Fig1E}
	\end{figure}
\end{center}

\newpage

\begin{center}
	\begin{figure}[!h]
		\includegraphics[width=0.8\linewidth]{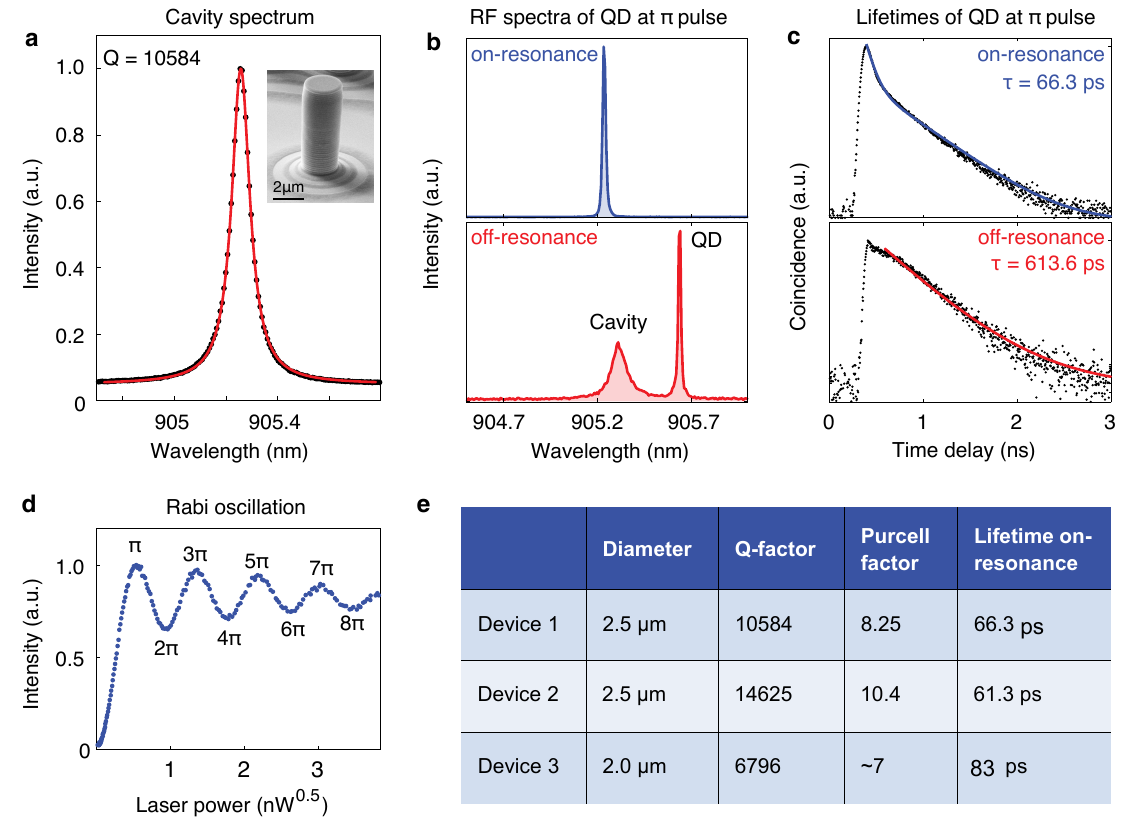}
		\caption{\textbf{Experimental parameters of the investigated devices.} (a) Cavity spectrum of device 1 under high-power above-band excitation. A Q-factor of 10584 is acquired by fitting the data with a Lorentz function. Insert: scanning electron microscope (SEM) image of device 1 with a diameter of 2.5 $\rm \mu m$. (b) $\pi$-pulse RF spectra of the QD in device 1 when QD is on/off-resonance with the cavity mode. (c) Corresponding on/off-resonance lifetimes of the QD at $\pi$ pulse. (d) Rabi oscillation of device 1 extracted from Fig.~\ref{fig:Fig2}(b), (e) Parameters of the devices investigated in the main text. }
		\label{fig:Fig2E}
	\end{figure}
\end{center}

\newpage

\begin{center}
	\begin{figure}[!h]
		\includegraphics[width=0.8\linewidth]{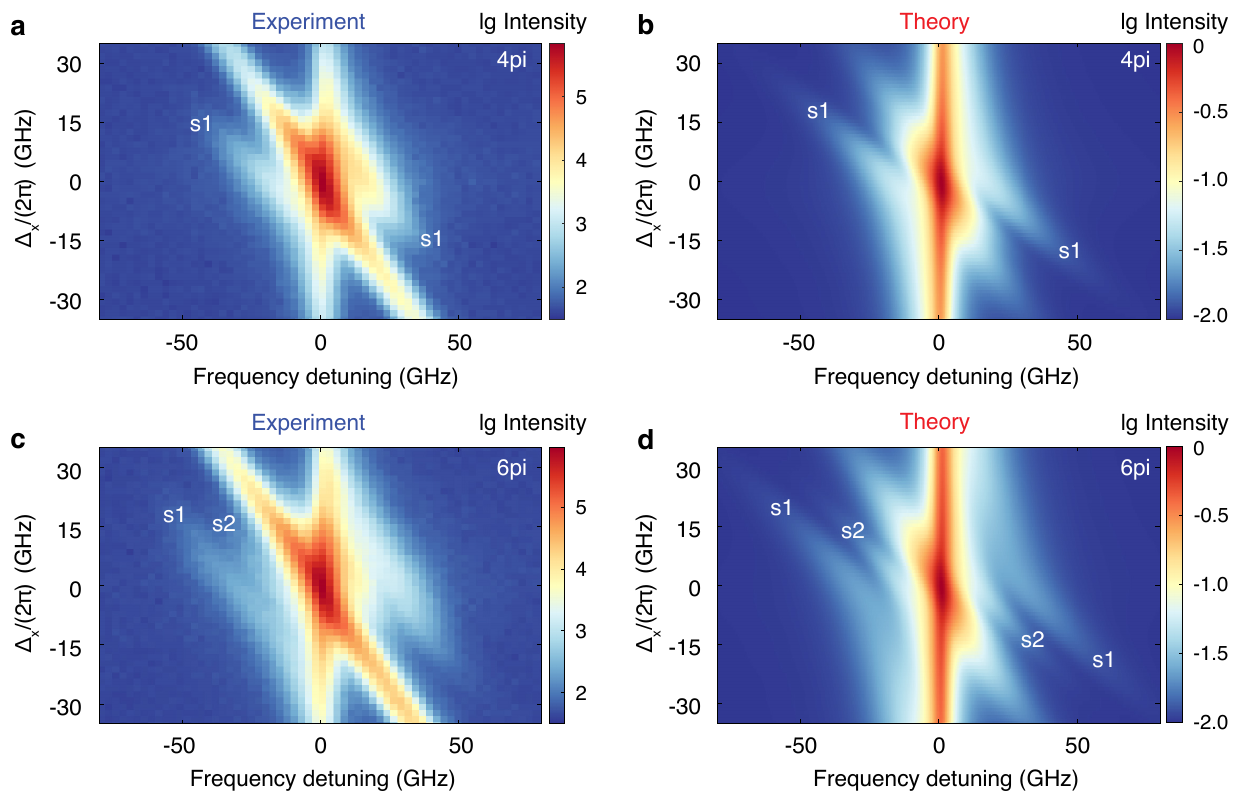}
		\caption{\textbf{Laser detuning dependent RF spectra.} (a) Measured laser detuning dependent dynamic RF spectra of device 1 at 4$\pi$ pulse area. (b) Calculated dynamic RF at 4$\pi$ pulse area with varied $\Delta_x$. (c) Measured laser detuning dependent dynamic RF of device 1 at 6$\pi$ pulse area. (b) Calculated dynamic RF at 6$\pi$ pulse area with varied $\Delta_x$. The pulse width is kept for 54 ps for this experiment.}
		\label{fig:Fig3E}
	\end{figure}
\end{center}

\newpage

\begin{center}
	\begin{figure}[!h]
		\includegraphics[width=0.9\linewidth]{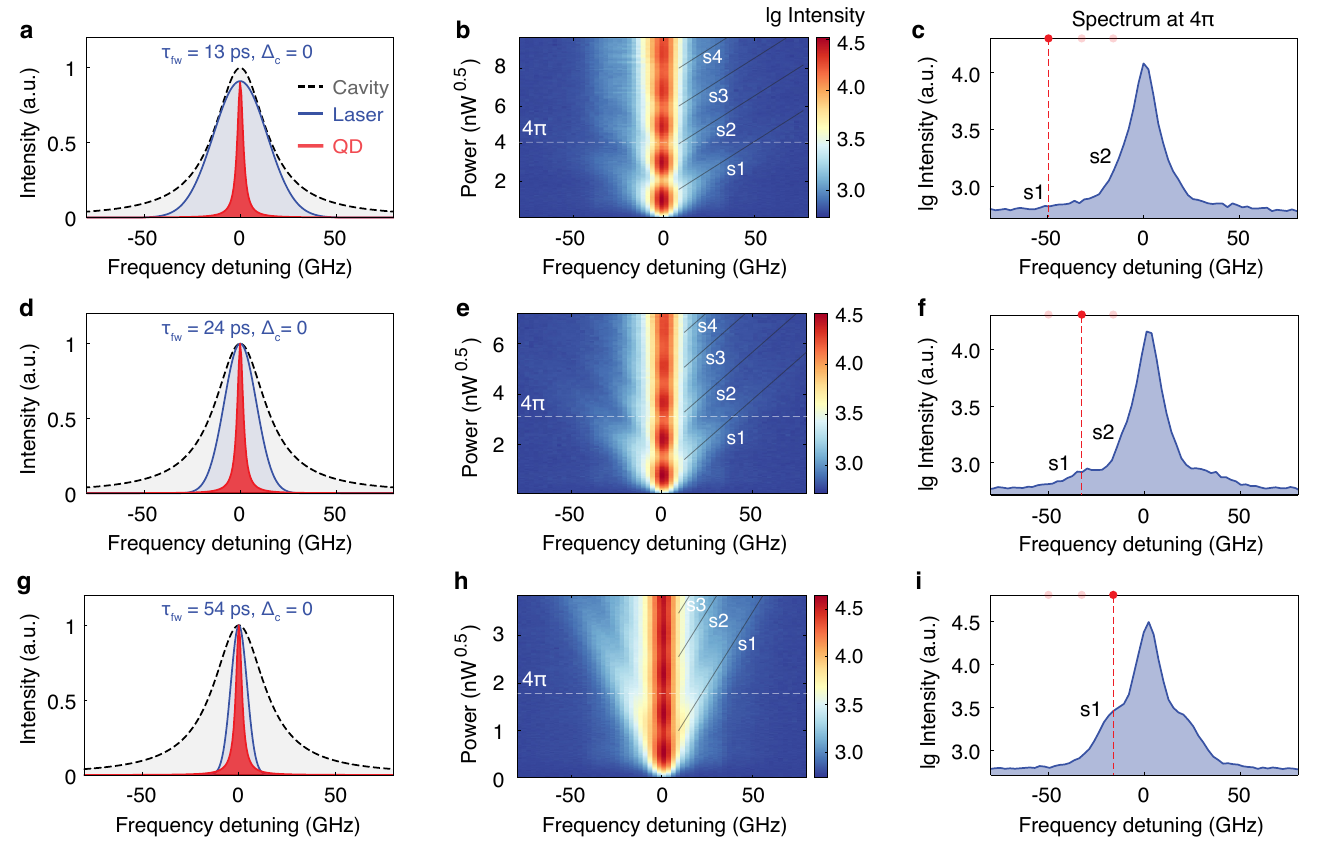}
		\caption{\textbf{Dynamic RF with varied pulse width $\tau_{\rm{fw}}$ (FWHM of the laser pulse) under the condition of $\Delta_c  = 0$ for device 1 at 14 K.} (a) Excitation condition: $\Delta_x=0$, $\tau_{\rm{fw}} = 13~\rm ps$. (b) Contour plots of the measured RF spectra [under the condition in (a)] as a function of driving pulse amplitude. The side peaks are labeled with s1 to s2. (c) Log-scale RF spectra [under the condition in (a)] at 4$\pi$ pulse areas. The position of side peak 1 is pointed out with the dashed line in red. (d) Excitation condition: $\Delta_x=0$, $\tau_{\rm{fw}} = 24~\rm ps$. (e) Contour plots of the measured RF spectra [under the condition in (d)] as a function of driving pulse amplitude. (f) Log-scale RF spectra [under a condition in (d)] at 4$\pi$ pulse areas. (g) Excitation condition: $\Delta_x=0$, $\tau_{\rm{fw}} = 54~\rm ps$. (h) Contour plots of the measured RF spectra [under the condition in (g)] as a function of driving pulse amplitude. (i) Log-scale RF spectra [under the condition in (g)] at 4$\pi$ pulse areas. The side peaks get close to each other and move toward the central peak as the pulse width increases.}
		\label{fig:Fig4E}
	\end{figure}
\end{center}

\newpage

\begin{center}
	\begin{figure}[!h]
		\includegraphics[width=\linewidth]{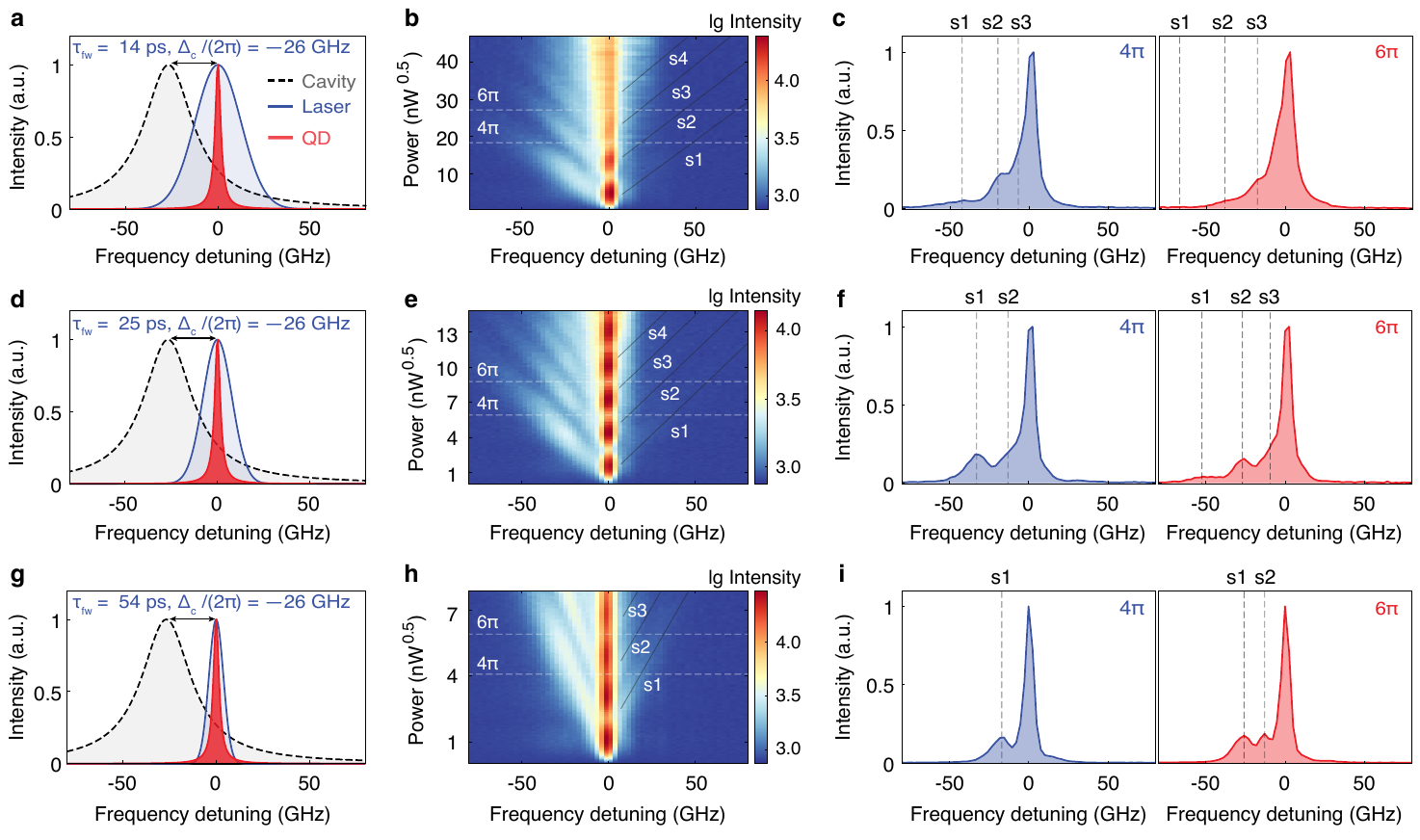}
		\caption{\textbf{Dynamic RF with varied pulse width $\tau_{\rm{fw}}$ under the condition of $\Delta_c/(2\pi)  = -26$ GHz for device 1 at 4 K.} (a) Excitation condition: $\Delta_x=0$, $\tau_{\rm{fw}} = 14~\rm ps$. (b) Contour plots of the measured RF spectra [under the condition in (a)] as a function of driving pulse amplitude. The side peaks are labeled with s1 to s3.  (c) RF spectra [under the condition in (a)] at 4$\pi$ and 6$\pi$ pulse areas. (d) Excitation condition: $\Delta_x=0$, $\tau_{\rm{fw}} = 25~\rm ps$. (e) Contour plots of the measured RF spectra [under the condition in (d)] as a function of driving pulse amplitude.  (f) RF spectra [under a condition in (d)] at 4$\pi$ and 6$\pi$ pulse areas. (g) Excitation condition: $\Delta_x=0$, $\tau_{\rm{fw}} = 54~\rm ps$. (h) Contour plots of the measured RF spectra [under the condition in (g)] as a function of driving pulse amplitude. (i) RF spectra [under a condition in (g)] at 4$\pi$ and 6$\pi$ pulse areas. Because of the enhancement of the cavity mode, the intensity of the peaks on the left side is comparable with that of the central peak. The side peaks are clearly observable even in the linear scale.}
		\label{fig:Fig5E}
	\end{figure}
\end{center}

\newpage
\begin{center}
	\begin{figure}[!h]
		\includegraphics[width=1.0\linewidth]{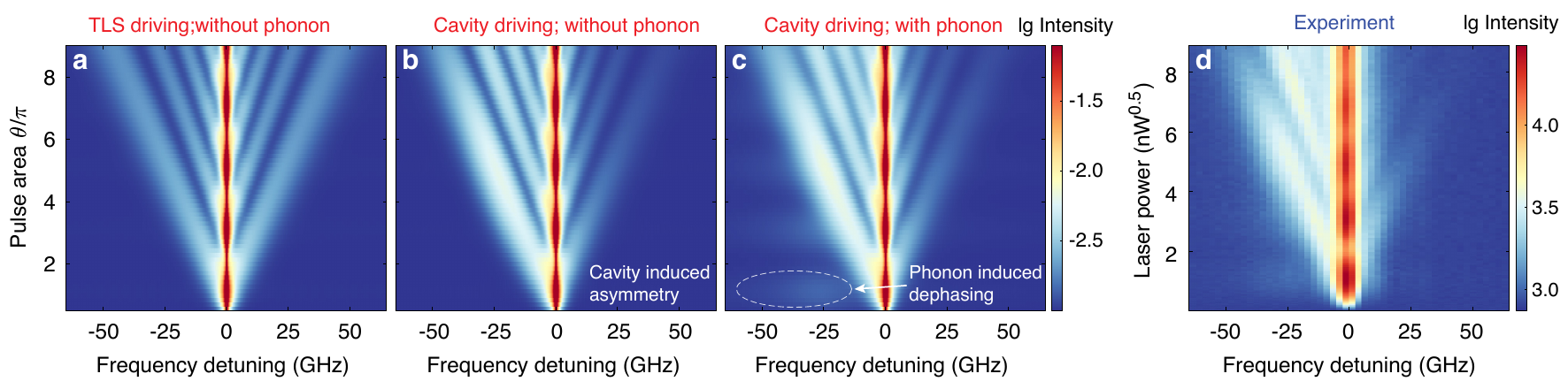}
		\caption{\textbf{Comparison of different models for reproducing the experimental result in Fig.~\ref{fig:Fig3}(c).} (a) The TLS is driven directly by the laser with $H_{\rm drive}(t) = \frac{\Omega(t)}{2}(\sigma^{+}+\sigma^{-})$, we use a naive bad-cavity model to eliminate the cavity mode, and phonons are not considered. (b) A full quantum model with a quantized cavity mode is adopted, with $H_{\rm drive}(t) = \frac{\Omega(t)}{2}(a+a^{\dagger})$, and phonons are not considered. The intensity asymmetry of the side peaks due to cavity filtering is reproduced, which agrees well with the experiment. (c) The cavity is driven with $H_{\rm drive}(t) = \frac{\Omega(t)}{2}(a+a^{\dagger})$, and phonons are included. 
        This full model gives rise to additional phonon-induced features, also visible in the experiment, as highlighted by the white dotted line. (d) Contour plots of the measured RF spectra as a function of drive pulse amplitude with a pulse width of 54 ps under the condition of $\Delta_x=0$ and $\Delta_c/(2\pi) =-26$ GHz at 4 K.}
		\label{fig:Fig6E}
	\end{figure}
\end{center}

\newpage
\begin{center}
	\begin{figure}[!h]
		\includegraphics[width=0.75\linewidth]{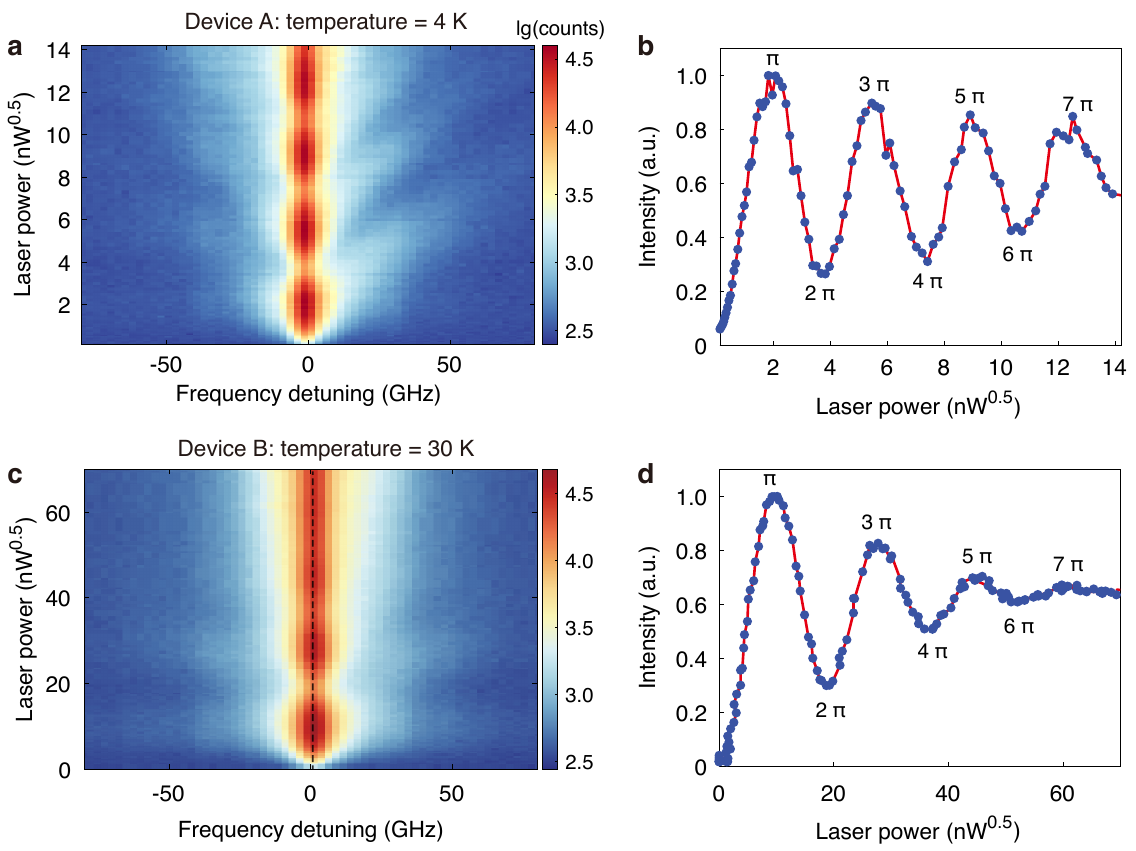}
		\caption{\textbf{Influence of phonons on dynamic RF spectra.}  Two additional devices (A and B) at the same sample with Q factors of $\sim$7000, Purcell factors of $\sim$7, and lifetimes of $\sim$80~ps. For device A, QD is in resonance with cavity mode at the temperature of 4 K, and for device B, QD is in resonance with cavity mode at the temperature of 30K. (a) Contour plot of the measured RF spectra as a function of driving pulse amplitude for device A. (b) Rabi oscillation of the main peak in (a). (c) Contours plot of the measured RF spectra as a function of driving pulse amplitude for device B. (d) Rabi oscillation of the main peak in (c). }
		\label{fig:Fig7E}
	\end{figure}
\end{center}

\end{document}